\documentclass[12pt]{article} 
\usepackage{graphicx} 
 
\newsavebox{\sboxpubnumber} 
\newsavebox{\sboxpubdate} 
\newcommand{\pubdate}[1]{\begin{lrbox}{\sboxpubdate}{#1}\end{lrbox}} 
\newcommand{\pubnumber}[1]{\begin{lrbox}{\sboxpubnumber}%
{\begin{tabular}{l} #1 \\ 
				 \usebox{\sboxpubdate} 
				 \end{tabular}} 
                           \end{lrbox} 
                           \pubblock} 
\newcommand{\Title}[1]{\begin{center} {\Large #1 } \end{center}} 
\newcommand{\Author}[1]{\begin{center}{ \sc #1} \end{center}} 
\newcommand{\Address}[1]{\begin{center}{ \it #1} \end{center}} 
 
\newcommand{\pubblock}{\rightline{ 
			\usebox{\sboxpubnumber}}} 
\newenvironment{Abstract}{\begin{quotation}  }{\end{quotation}} 
\newenvironment{Presented}{\begin{quotation} \begin{center} 
             PRESENTED AT\end{center}\bigskip 
      \begin{center}\begin{large}}{\end{large}\end{center} 
      \end{quotation}} 
 
\newcommand{\ba}{\begin{eqnarray}}

\newcommand{\ea}{\end{eqnarray}}
\newcommand{\rmi}[1]{{\mbox{\scriptsize #1}}}
\def\lsi{\raise0.3ex\hbox{$<$\kern-0.75em\raise-1.1ex\hbox{$\sim$}}}
\def\gsi{\raise0.3ex\hbox{$>$\kern-0.75em\raise-1.1ex\hbox{$\sim$}}}
\newcommand{\lsim}{\mathop{\lsi}}
\newcommand{\gsim}{\mathop{\gsi}}
\newcommand{\la}[1]{\label{#1}}
\newcommand{\be}{\begin{equation}}
\newcommand{\ee}{\end{equation}}
\newcommand{\fig}{Fig.~}
\newcommand{\eq}{Eq.~}

\newcommand{\nr}[1]{(\ref{#1})}
\newcommand{\tr}{{\rm Tr\,}}

\newcommand{\fr}[2]{{\frac{#1}{#2}\,}}

\newcommand{\tinymsbar}{{\overline{\mbox{\tiny\rm{MS}}}}}

\begin{document} 
%%%%%%%%%%%%%%%%%%%%%%%%%%%%%%%%%%%%%%%%%%%%%%%%%%%%%%%%%%%%%%%%%%%%%%%% 
%%%%%%%%%%%%%%%%%%%%%%%%%%%%%%%%%%%%%%%%%%%%%%%%%%%%%%%%%%%%%%%%%%%%%%%% 
\begin{titlepage} 
\pubdate{November 2001}                    %fill in the date 
\pubnumber{CERN-TH/2001-336\\ 
           hep-ph/0111349}                 %preprint number(s) 
 
\vfill 
\Title{Thermal phase transitions in cosmology} 
\vfill 
\Author{M.~Laine\footnote{Work partly supported by the TMR network {\em Finite
  Tempe\-ra\-ture}  {\em Phase Transitions in Particle Physics}, EU Contract
  No.\ FMRX-CT97-0122.}}
\Address{Theory Division, CERN, CH-1211 Geneva 23, Switzerland}
\vfill 
\begin{Abstract} 
We review briefly the current status of thermal phase transitions 
within the Standard Model and its simplest extensions. We start with 
an update on QCD thermodynamics, then discuss the electroweak phase transition,
particularly in supersymmetric extensions of the Standard Model, and 
end with a few remarks on the cosmological constraints that thermal phase 
transitions might impose on even higher scale particle physics.
\end{Abstract} 
\vfill 
\begin{Presented} 
    COSMO-01 \\ 
    Rovaniemi, Finland, \\ 
    August 29 -- September 4, 2001 
\end{Presented} 
\vfill 
\end{titlepage} 
\def\thefootnote{\fnsymbol{footnote}} 
\setcounter{footnote}{0} 
 
%%%%%%%%%%%%%%%%%%%%%%%%%%%% SECTION %%%%%%%%%%%%%%%%%%%%%%%%%%%%%%%%%%% 
\section{Introduction} 

As is well known, 
the microscopic energy scales of quantum mechanics 
and the macroscopic properties of our present Universe
are intimately connected. For 
instance, the ${\cal O}($eV) energy scale of atomic physics
manifests itself through the existence of the cosmic microwave 
background radiation, and the ${\cal O}($MeV) scale of nuclear 
physics through the primordial origin of light element abundances. 
The connection might of course extend even much further on:  
the small fluctuations observed in the microwave background 
could have been produced during an early period of inflation, 
which could well be a manifestation of particle physics at, say, 
the ${\cal O}(10^{15}$ GeV) scale of grand unification. 

With this background, it is natural to expect that 
the main scales of experimentally accessible
particle physics should also leave their marks 
in cosmology. Could not the ${\cal O}($GeV$)$ scale 
of QCD, and the ${\cal O}($TeV$)$ scale of the electroweak (EW) 
theory, be related to such cosmological remnants as extragalactic
magnetic fields, or baryon asymmetry? 

Taking a closer look, 
it turns out that quite non-trivial conditions have to be met
for the latter connections to exist.
Among the biggest challenges are: 

{\bf 1.}\ Generic QCD and EW interactions are so strong
that it is difficult to deviate from thermal equilibrium, which
is a necessary requirement for any cosmological remnant to emerge. 
Denoting by $n$ a particle density, $\sigma$ a cross section, 
$v$ an average velocity, $\alpha=g^2/(4\pi)$ the coupling,  
$T$ the temperature, and $m_\rmi{Pl} \sim 10^{19}$ GeV
the Planck mass, interactions fall out of equilibrium provided that 
\be
\tau \equiv \frac{1}{n \sigma v} \sim \frac{1}{\alpha^2 T} \gg 
t \equiv \frac{m_\rmi{Pl}}{T^2}, 
\ee
which leads to $T > \alpha^2 m_\rmi{Pl} \sim 10^{15}$ GeV. 
Thus, they can fall out of equilibrium only at temperatures 
where anyway the QCD and EW theories merge to a grand unified theory. 

There is one major exception to the conclusion just drawn: 
a theory possessing a {\em first order phase transition}
falls out of equilibrium, even if microscopic 
interactions are strong enough to be in local thermal equilibrium 
above and below the transition point.

{\bf 2.}\ Spatial fluctuations (relevant to remnants such as magnetic 
fields) are produced only on very small length scales.
Indeed, the horizon of the moment when $T\sim 1$ GeV
corresponds today to about 1 light year, and that of $T\sim 1$ TeV to about 
1 astronomical unit. Fluctuations can effectively only be produced
on scales smaller than these, which leads to miniscule numbers  
with respect to intergalactic distances. 

Again, there is one conceivable way of avoiding 
the problem: magnetohydrodynamic evolution is very non-linear
and could potentially transfer energy to large 
length scales more rapidly than comoving expansion~\cite{b}.

The subject of this talk is the analysis of the first of the problems
mentioned, the existence of first order phase transitions in QCD 
and in various
versions of the EW theory. The actual generation of cosmological 
remnants has been treated in other talks at this conference~\cite{jc,mg}, 
and will only be touched upon very briefly here.

%%%%%%%%%%%%%%%%%%%%%%%%%%%% SECTION %%%%%%%%%%%%%%%%%%%%%%%%%%%%%%%%%%% 
\section{QCD thermodynamics}

The theory of strong interactions, QCD, is expected to undergo
a phase transition at a temperature of the order 
of $\Lambda_\tinymsbar$. The transition is said to be related 
to deconfinement and chiral symmetry restoration. However, 
neither of these are rigorous concepts for physical quark masses,  
and therefore there might just as well be a smooth gradual 
change of the properties of the system, 
instead of an actual singularity. 

In any case, a smooth change is not what one naively expects. 
Indeed, counting the ``free'' degrees of freedom $g_*$ in the ``confined'' 
and ``deconfined'' phases, one finds a considerable change, 
suggesting perhaps a strong transition: 
\ba
 g_{*}({T< T_c}) & = &  
\mbox{(pions)} + ... = 17.25, \\
g_{*}({T> T_c}) & = &
\mbox{(gluons)} +\fr78\mbox{(light quarks)} + ... = 51.25.
\ea 
However, in practice, particles are not free but strongly interacting. 
This leads to the fact that the properties of the QCD
phase transition can, on a quantitative level, only be studied systematically
with four-dimensional (4d) finite temperature lattice simulations. 

It turns out, furthermore, that such lattice simulations are 
very demanding. The reason is that
the precise characteristics of the transition, such as its order, 
depend strongly on the symmetries of the system, which
are in turn determined by the quark masses (for a review, see~\cite{fk}). 
But chiral quarks, as light as they are in Nature, are difficult to 
fit on the lattices available in practice. 
Thus it is believed that the properties of the system 
do change at a temperature $T_c \sim 170$ MeV~\cite{tc}, 
but whether the change is smooth or, in the large volume 
limit, discontinuous, remains open. 

\begin{figure}[htb] 
    \centering 
    \includegraphics[height=6cm]{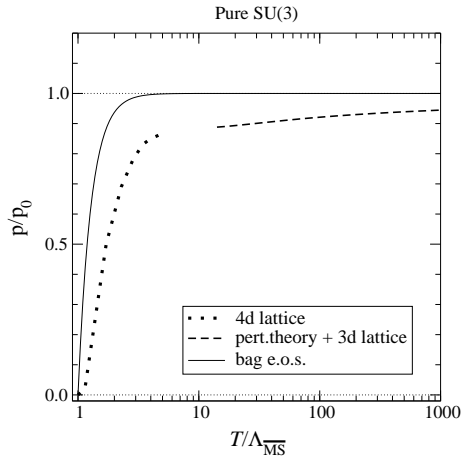} 
    \caption{The pressure of pure SU(3) gauge theory, compared with 
             the free Stefan-Boltzmann value $p_0 = (\pi^2/45)(N_c^2-1) T^4$, 
             as a function of $T/{\Lambda_\tinymsbar}$, where 
             $\Lambda_\tinymsbar$ is the scale parameter.
             The transition takes place at $T\approx \Lambda_\tinymsbar$.
             The 4d lattice results are from~\cite{boyd}, and the curve 
             labelled perturbation theory + 3d lattice, from~\cite{a0cond}.
             The results are compared with the ``bag'' equation
             of state, $p_\rmi{bag}(T) \equiv p_0(T) - p_0(T_c)$.} 
    \label{fig:qcd} 
\end{figure}

What is known much better is the behaviour of
various thermodynamical quantities, such as the pressure, at 
temperatures above the critical, $T \gsim T_c$. 
Indeed, there the inclusion of quarks~\cite{nf2}
does not appear to change the result qualitatively from the idealised case
of pure SU(3)~\cite{boyd}. The general pattern,
illustrated in~\fig\ref{fig:qcd}, is that  
the pressure is small at $T\sim T_c$, rises rapidly at 
$T\sim (1...2) T_c$, and then levels off, approaching
the ideal gas limit very slowly, with a deviation of 10...15\% 
for a long while.  

%%%%%%%%%%%%%%%%%%%%%%%%%%%% FIGURE %%%%%%%%%%%%%%%%%%%%%%%%%%%%%%%%%%%% 
%%%%%%%%%%%%%%%%%%%%%%%%%%%%%%%%%%%%%%%%%%%%%%%%%%%%%%%%%%%%%%%%%%%%%%%% 

This noticeable deviation from 
ideal gas thermodynamics has some implications. 
According to the Einstein equations, the temperature
in the Universe evolves as
\be
\frac{dT}{dt} = - \frac{\sqrt{24\pi}}{m_\rmi{Pl}}
\frac{\sqrt{e(T)}}{d\ln s(T)/dT}. 
\ee
Here $s=p'(T), e=T^2 (p(T)/T)'$. Using an equation of state
as in~\fig\ref{fig:qcd}, 
the Universe cools slower than a free gas~\cite{ikkl}, 
and the sound velocity $v_s^2 = p'(T)/e'(T)$, characterising
hydrodynamic fluctuations, dips around the transition point. 
These facts do result in a non-trivial  fluctuation
spectrum, even if long-lasting
consequences might not remain. 

%%%%%%%%%%%%%%%%%%%%%%%%%%%% SECTION %%%%%%%%%%%%%%%%%%%%%%%%%%%%%%%%%%% 
\section{EW phase transition} 

At temperatures of the order of $T_c \sim m_H/g$, where 
$m_H$ is the Higgs mass and $g$ is the weak coupling constant, 
the electroweak gauge symmetry gets restored~\cite{early}.
This may have led to an important consequence,  
the existence of a matter--antimatter (baryon) asymmetry~\cite{krs}
(for reviews, see~\cite{rs,mq,jc}). Again, a deviation from equilibrium
through a first order phase transition is a necessary requirement. 
More quantitatively, a 1-loop saddle point computation~\cite{bj}
as well as a non-perturbative evaluation~\cite{moore_broken}
show that the discontinuity in the Higgs expectation value
across the phase transition, $\Delta v/T$ (in, say, the Landau gauge) 
should exceed unity, $\Delta v/T \gsim 1.0$.

In contrast to the QCD case, many properties
of the EW transition can be addressed in perturbation theory.
This is simply because the Higgs mechanism itself is perturbative. 
Eventually perturbation theory breaks down, though: at finite temperatures 
the largest loop expansion parameter can be estimated to be~\cite{nonpert} 
\be
\epsilon \sim \frac{g^2 T }{\pi m}, \la{epsilon}
\ee
where $m$ is some mass scale. Thus light
excitations, $m\lsim g^2T$, always present at the phase 
transition point, lead to an infrared problem~\cite{gpy}. 

The expansions parameters related to 
heavy excitations, on the other hand, 
such as the non-zero Matsubara modes of finite temperature field theory
with $m\sim \pi T$, are small. This allows for 
an essential simplification of the non-perturbative treatment:
one can integrate out massive modes perturbatively 
and study only the dynamics of the light modes non-perturbatively~\cite{krs2}.
In the first step, the original 4d theory
reduces to a three-dimensional (3d) effective one, 
in a process called dimensional reduction~\cite{dr}. 
For studying the EW
phase transition in a weakly coupled 
theory ($m_H\lsim 250$ GeV), 
this approach works with a practical accuracy at the per cent 
level, from the point of view of both dimensional 
reduction~\cite{generic}--\cite{gr2} and numerical 
simulations~\cite{nonpert,su2u1,owe}.

In the case of the Standard Model and many of its
extensions~\cite{generic}, the only non-perturbative
infrared modes are a Higgs doublet $\phi$ and 
the spatial SU(2) and U(1) 
gauge fields, with field strength tensors $F_{ij},B_{ij}$
and gauge couplings $g_3,g_3'$. The Lagrangian is
\ba
{\cal L}_{\rm 3d} & = & 
{1\over2}\tr F_{ij}^2 +{1\over4} B_{ij}^2+
(D_i\phi)^\dagger D_i\phi+m_3^2\phi^\dagger\phi+
\lambda_3(\phi^\dagger\phi)^2.
\label{lagr}
\ea
All knowledge about the 
physical zero temperature parameters, as well as about 
the temperature, 
is encoded in the expressions for the effective couplings 
$m_3^2,\lambda_3,g_3^2,g_3'^2$. 

The properties of the phase transition now depend
on the effective couplings in~\eq\nr{lagr}. 
In particular, since the gauge couplings are fixed
and the mass parameter $m_3^2$ is to be tuned to the
phase transition point, such properties are determined 
by the single dimensionless ratio
${\lambda_3}/{g_3^2}$~\cite{generic}.
To get a feeling of this dependence, let 
us first apply 1-loop perturbation theory. Ignoring the 
tiny corrections from $g_3'^2/g_3^2$, we find 
a first order transition with the discontinuity
\be
\Delta \frac{v}{gT} = \frac{1}{8\pi} 
\frac{g_3^2}{\lambda_3}\biggl[1 + 
{\cal O}\Bigl(\frac{\lambda_3}{g_3^2}\Bigr)\biggr].
\label{disc}
\ee
Thus, the transition weakens for large $\lambda_3/g_3^2$. 
In reality, the line of first order transitions even ends completely, 
at $\lambda_3/g_3^2 = 0.0983(15)$~\cite{endpoint}. 

%%%%%%%%%%%%%%%%%%%%%%%%%%%%%%%%% FIGURE %%%%%%%%%%%%%%%%%%%%%%%%%%%%%%%%%%
\begin{figure}[t!]
    \centering 
    \includegraphics[height=6cm]{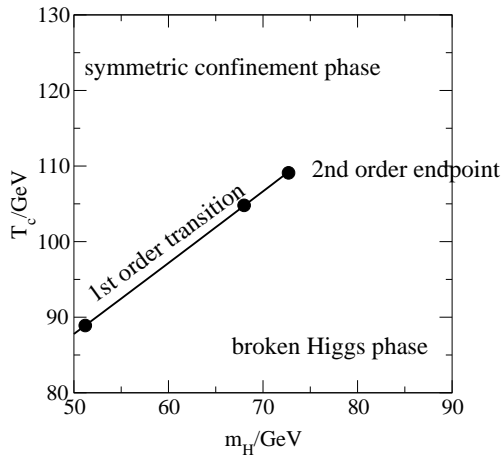} 
    \caption[a]{The phase diagram of the physical Standard Model. The blobs
    indicate simulation points (with error bars)~\cite{su2u1,endpoint}, 
    and the solid curve is a fit through them.}

\label{fig:sm}
\end{figure}
%%%%%%%%%%%%%%%%%%%%%%%%%%%%%%%%%%%%%%%%%%%%%%%%%%%%%%%%%%%%%%%%%%%%%%%%%%%

\subsection{Standard Model}

To be now more specific, let us consider the Standard Model. Then 
\be
\frac{\lambda_3}{g_3^2} \approx {1\over 8} \frac{m_H^2}{m_W^2} + 
{\cal O} \biggl( \frac{g^2}{(4\pi)^2} \frac{m_\rmi{top}^4}{m_W^4}\biggr).
\label{efflam}
\ee
Applying this (with the actual 1-loop corrections properly included)
to the lattice results of~\cite{su2u1,endpoint} leads to 
Fig.~\ref{fig:sm}. We find that the endpoint location
$\lambda_3/g_3^2 = 0.0983(15)$ corresponds in physical units to
$m_{H,c}= 72.3(7)$ GeV, $T_c = 109.2(8)$ GeV~\cite{endpoint}.

As pointed out above, 
for baryogenesis we would need a first order phase transition, and 
even a strong one, $\Delta v/T\gsim 1$. We have learned that 
a first order transition only exists for $m_H < 72$ GeV. 
Furthermore, it has in fact $\Delta v/T \lsim  1.0$ down to
$m_H\sim 10$ GeV~\cite{ae,nonpert}!
Thus, we observe that experimentally allowed Higgs masses 
$m_H \gsim 115$ GeV~\cite{lep} are very far from allowing
for electroweak baryogenesis. 

It appears that primordial magnetic fields present at the time
of the EW transition can strengthen the transition quite significantly,
but not enough to change the conclusions~\cite{bext}, though
they are associated with other intriguing phenomena~\cite{allor}.

\subsection{Supersymmetric extensions of the Standard Model}

How should the Higgs sector be modified in order to change the pattern above?
As the strength of the transition is determined by the scalar 
self-coupling, cf.\ Eq.~(\ref{disc}),
we apparently need some new degree of freedom, which can decrease 
the effective $\lambda_3$ 
by ${\cal O}(100\%)$. It turns out that 
to get such a large correction, 
we need a bosonic Matsubara zero mode with an expansion 
parameter as in Eq.~(\ref{epsilon}). 
A simple perturbative 1-loop 
computation shows that at least the sign of such 
loop corrections is the correct one, 
negative. But to have an effect
of ${\cal O}(100\%)$, we need
$\epsilon \sim 1$, so that $m \sim g^2 T/\pi$.
That is, the degree of freedom should itself be non-perturbative!

In the MSSM, natural candidates for such degrees of freedom
are squarks and sleptons. Then there are thermal corrections 
in their effective mass parameters, appearing as 
$m_3^2 \sim m_\rmi{4d}^2 + g^2T^2$. Thus, to get a total 
outcome of order $(g^2 T )^2$, a cancellation must take place,  
which requires a negative mass parameter $m_\rmi{4d}^2$. 
At zero temperature, the physical mass is then roughly
$m_\rmi{phys}^2 \lsim  m_\rmi{4d}^2 + m_\rmi{top}^2 \lsim m_\rmi{top}^2$. 
In order for such a relatively light degree of freedom not to have shown
up so far in precision observables, 
it should be an SU(2) singlet. Since it should also couple
strongly to the Higgs, we should choose stops.

%%%%%%%%%%%%%%%%%%%%%%%%%%%%%%%%% FIGURE %%%%%%%%%%%%%%%%%%%%%%%%%%%%%%%%%%
\begin{figure}[t!]
    \centering 
    \includegraphics[height=6cm]{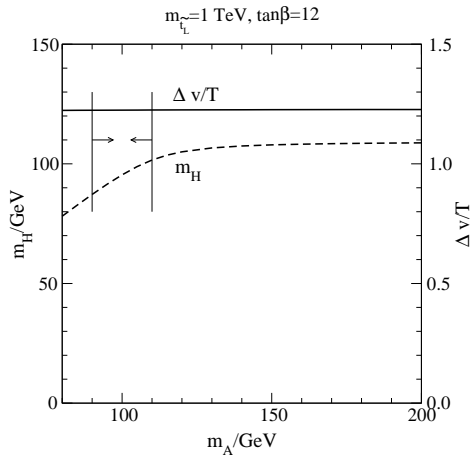} 
    \caption[a]{Examples of values 
    (Higgs mass $m_H$, mass parameter $m_A$), leading to a large
    $\Delta v/T$ according to 2-loop perturbation theory, for an optimal
    choice of $m_{\tilde t_R}$~(from~\cite{mssmsim}). The region indicated
    with arrows (widening for increasing $m_{\tilde t_L}$~\cite{bjls,mssmsim})
    appears to be within experimental constraints,  
    even for the most stringent ``no mixing'' scenario~\cite{lep}.}

\label{fig:extensions}
\end{figure}
%%%%%%%%%%%%%%%%%%%%%%%%%%%%%%%%%%%%%%%%%%%%%%%%%%%%%%%%%%%%%%%%%%%%%%%%%%%

Now, the stops come left- and right-handed, $\tilde t_L, \tilde t_R$ 
(these states can also mix, but for lack of space we ignore this here).
The requirement of having a small violation of the 
electroweak precision observables means that the weakly 
interacting left-handed one cannot be light, $m_{\tilde t_L}\gg m_\rmi{top}$.
This also increases the Higgs mass
(see, e.g.,~\cite{erz}), 
\be
m_H^2 \sim  m_Z^2\cos^2\! 2\beta +  
\frac{3 g^2}{8 \pi^2} \frac{m_\rmi{top}^4}{m_W^2}
\ln\frac{m_{\tilde{t}_R}m_{\tilde{t}_L}}{m_\rmi{top}^2}.
\label{mH}
\ee
On the contrary, the SU(2) singlet stop $\tilde t_R$ can 
be ``light'', and serve as the desired new degree of 
freedom~\cite{tr_early}. (Then, of course, we lose half of the correction
to $m_H$ in Eq.~(\ref{mH}), but this is the price to pay.)
It should not be so light that the stop 
direction gets broken before the phase transition, 
though, because then one cannot get back to the EW minimum 
afterwards~\cite{cms}.
Another example of a viable light scalar degree of 
freedom is the complex gauge singlet of the NMSSM~\cite{hsch}. 

Under the conditions described above, 
resummed 2-loop perturbation theory 
indicates that the EW phase
transition can indeed be strong enough 
for baryogenesis~\cite{bjls}.
To check the reliability of perturbation theory, a 3d effective
theory (a generalisation of~\eq\nr{lagr}) has again
been constructed and studied with simulations~\cite{mssmsim}, 
with the encouraging outcome that, in fact, 
for a light right-handed stop and heavy Higgs, the 2-loop estimates 
are reliable and even somewhat conservative.
This is in strong contrast to the case of the 
Standard Model at realistic Higgs masses. 

Next, we must ask 
whether this parameter region is indeed in agreement 
with all experimental data. The most important constraint 
comes from the lower bound on the Higgs mass~\cite{lep}.
There is a parameter in the MSSM, $m_A$, 
which determines whether the Higgs sector resembles that in the 
Standard Model ($m_A \gsim 120$~GeV) or not ($m_A \lsim 120$~GeV).
In the latter case, the experimental lower bound is relaxed~\cite{lep},
and since the transition needs not always get significantly weaker
(see Fig.~\ref{fig:extensions}), this case is acceptable.  
If $m_A$ is larger, then the left-handed stop should be quite heavy,
$m_{\tilde t_L}\gsim 2$ TeV, 
in order to increase the Higgs mass towards the Standard Model
value $\gsim 115$~GeV (cf.~\eq\nr{mH}). 
Allowing for significant mixing in the stop
mass matrix relaxes the Higgs mass bounds quite considerably~\cite{lep}, 
although it also tends to weaken the transition
somewhat~\cite{bjls,mssmsim}.

%%%%%%%%%%%%%%%%%%%%%%%%%%%% SECTION %%%%%%%%%%%%%%%%%%%%%%%%%%%%%%%%%%% 
\section{Phase transitions at higher scales?}
 
Let us end by considering hypothetical phase transitions at scales
even higher than the EW one, possibly related to unification. 
To discuss this self-consistently, it must be assumed that
inflation and unification are not related, but inflation takes 
place earlier on, maybe at the Planck scale. Then, in clear contrast
to the cases considered so far, thermal phase transitions rather
generically do tend to produce cosmological remnants, whose 
non-observation allows to place constraints on possible 
unification models, or on cosmology. 
We are here referring to topological defects, 
such as domain walls, cosmic strings, and monopoles~\cite{defect}. 
As an example, let us make some more specific comments on the first ones. 

The reason why there is a strong constraint is that 
the energy density $\delta\rho$ related to domain walls 
decays much more slowly than radiation. 
It would lead to $\delta \rho/\rho \gsim 10^{-5}$ at photon 
decoupling, if  the domain wall surface energy density is
$\sigma \gsim (1\mbox{ MeV})^3$. This exceeds the fluctuations
observed in the cosmic microwave background. Thus {\em any} new
theories leading to thermal domain wall production are excluded~\cite{z}!

As timely examples, let us consider 
models with compact extra dimensions. In case there
are gauge fields
living in the bulk~\cite{ia} but no fundamental matter, 
one generically gets Z($N_c$) domain walls~\cite{bk}. 
Thus, a class of such models
may be constrained by the thermal phase transitions
they would undergo in cosmology. It should also be mentioned 
that  the winding of a brane around the extra dimension
may lead to various types of cosmic strings and monopoles~\cite{dks}, 
potentially resulting in other constraints.

%%%%%%%%%%%%%%%%%%%%%%%%%%%% SECTION %%%%%%%%%%%%%%%%%%%%%%%%%%%%%%%%%%% 
\section{Conclusions} 

To summarise,
the QCD phase transition is
probably very weak, if there at all. However, 
the system is strongly interacting, and its thermodynamics deviates 
significantly from that of an ideal gas, which leads to a lengthy
period of non-standard expansion.  

In EW theories, there is in general no phase transition at all 
for realistic Higgs masses, unless there is also another light 
scalar degree of freedom, which plays an essential role in phase 
transition dynamics. 
For instance, the transition can be strong enough for baryogenesis 
in the MSSM if there is one very light ($m_{\tilde t_R} \lsim m_\rmi{top}$)
and one rather heavy stop. Either the heavy stop should be really heavy, 
$m_{\tilde t_L} \gsim 10\, m_\rmi{top}$, or the 
Higgs sector should contain at least 
two independent light particles, $m_A \lsim 120$ GeV, 
in order not to violate experimental constraints.
There is much more freedom in the NMSSM.

While little concrete can be said about theories
of unification scale physics, one can at least note that the possible  
overproduction of topological defects in phase transitions 
can place some constraints on model building and cosmology
(on top of many other constraints, of course). 
A classic example is the monopole problem, 
but one can also arrive at a domain wall problem in some 
models with extra dimensions. Such problems can be avoided if the 
reheating temperature after inflation is below the unification scale. 

Finally, let us stress that the topic of this talk has been phase 
transitions taking place in a system in local thermodynamical
equilibrium. In case of low-scale inflation ending at a reheating
temperature $T\sim {\cal O}$(TeV), 
it is also natural
to consider non-thermal phase transitions (see, e.g.,~\cite{others}), 
which lead to many new physics possibilities.
 
%%%%%%%%%%%%%%%%%%%%%%%%%%%% SECTION %%%%%%%%%%%%%%%%%%%%%%%%%%%%%%%%%%% 
%\Acknowledgements 
 
%%%%%%%%%%%%%%%%%%%%%%%%%%%%%%%%%%%%%%%%%%%%%%%%%%%%%%%%%%%%%%%%%%%%%%%% 
%%%%%%%%%%%%%%%%%%%%%%%%%% BIBLIOGRAPHY %%%%%%%%%%%%%%%%%%%%%%%%%%%%%%%%

\end{document}